\documentclass[preprint,showpacs,amsmath,amssymb,floatfix]{revtex4-1}

\usepackage{color}%OMPtion added by cyw eg \pagecolor{blue} \color{red}
\usepackage{graphicx}% Include figure files
\usepackage{subfig}
\usepackage{dcolumn}% Align table columns on decimal point
\usepackage{bm}% bold math
\usepackage{enumerate}
\usepackage{setspace}
\usepackage{listings}
\usepackage{xcolor}
\usepackage{tabularx,ragged2e,booktabs,caption}

\normalsize

\begin{document}

%\begin{CJK*}{GB}{} % Use default fonts from CJK (see below)

\title{Super-transition-array calculations for synthetic spectra
and opacity of high-density, high-temperature germanium
plasmas}

\author{Teck-Ghee.~Lee$^{a,*}$, W.~Jarrah$^b$, D.~Benredjem$^{b,\dag}$, J.-C.~Pain$^c$,  
M.~Busquet$^d$, M.~Klapisch$^e$, A.~J.~Schmitt$^a$, J.~W.~Bates$^a$, and J.~Giuliani$^a$}
%\email{teckghee.lee@nrl.navy.mil, $^{\dag}$djamel.benredjem@u-psud.fr}   
\affiliation{$^a$U.S. Naval Research Laboratory, Washington, D.C., USA}
\affiliation{$^b$Laboratoire Aim\'{e} Cotton (CNRS, Universit\'{e} Paris-Sud, ENS Paris-Saclay), 91405 Orsay, France}
\affiliation{$^c$CEA, DAM, DIF, F-91297 Arpajon Cedex, France}
\affiliation{$^d$Research Support Instruments, 4325 Forbes Blvd, Lanham, MD, USA}
\affiliation{$^e$Syntek Technologies Inc., 801 N Quincy St Ste 610, Arlington, VA, USA}

%\author{T.-G.~Lee}
%\email{teckghee.lee@nrl.navy.mil}
%\affiliation{U.S. Naval Research Laboratory, Washington, D.C., USA}

%\author{W.~Jarrah}
%\affiliation{Laboratoire Aim\'{e} Cotton (CNRS, Universit\'{e} Paris-Sud, ENS Paris-Saclay), 91405 Orsay, France}

%\author{D.~Benredjem}
%\email{djamel.benredjem@u-psud.fr}
%\affiliation{Laboratoire Aim\'{e} Cotton (CNRS, Universit\'{e} Paris-Sud, ENS Paris-Saclay), 91405 Orsay, France}

%\author{J.-C.~Pain}
%\affiliation{CEA, DAM, DIF, F-91297 Arpajon Cedex, France}

%\author{M.~Busquet}
%\affiliation{Research Support Instruments, 4325 Forbes Blvd, Lanham, MD, USA}

%\author{M.~Klapisch}
%\affiliation{Syntek Technologies Inc., 801 N Quincy St Ste 610, Arlington, VA, USA}

%\author{A.~J.~Schmitt}
%\affiliation{U.S. Naval Research Laboratory, Washington, D.C., USA}

%\author{J.~W.~Bates}
%\affiliation{U.S. Naval Research Laboratory, Washington, D.C., USA}

%\author{J.~Giuliani}
%\affiliation{U.S. Naval Research Laboratory, Washington, D.C., USA}
%\author{}
%\address{}

\makeatletter
\def\blfootnote{\gdef\@thefnmark{}\@footnotetext}
\makeatother
\blfootnote{$^{\ast}$teckghee.lee@nrl.navy.mil; $^{\dag}$djamel.benredjem@u-psud.fr.}
%\blfootnote{DISTRIBUTION A. Approved for public release: distribution unlimited.}


\def\bb#1{\hbox{\boldmath${#1}$}}
\def\pbar {\bar p}
%\setstretch{1.5}   %double spacing 

\begin{abstract}
The synthetic emission spectra and opacity of high-density, high-temperature
germanium (Z=32) plasma from super-transition-array (STA) 
calculations are presented. The viability of the STA model, which is based on 
a statistical superconfigurations accounting approach for calculating the atomic 
and radiative properties, is examined by comparing and contrasting its results 
against the available experimental data and other theoretical calculations. First, we focus on the 
emission data. To model the data, the Eulerian radiation-hydrodynamics code FastRad3D 
is used in conjunction with STA to obtain the STA-required inputs, 
namely, the time-dependent temperature and density profiles 
of the Ge plasmas. Consequently, we find that STA results fit the experimental spectrum 
reasonably well, reproducing the main spectral features of $2p-3d$, $2s-3p$, and $2p-4d$ 
transitions from the laser-heated germanium layer buried in plastic [High Energy Density Phys., 6 (2010) 105]. 
However, careful comparison between experimental and theoretical results 
in the photon-energy regions of $\sim$1.7 keV shows some degrees of disparity between the two. 
This may be due to the non-LTE effects and the presence of spatial gradients in the sample. 
Limitations of STA to model the experimental spectrum precisely is expected and 
underscoring the difficulty of the present attempts as the model assumed local thermodynamics 
equilibrium population dynamics. Second, we examine the STA calculated multi-frequency opacities 
for a broad range of Ge plasma conditions covering the L- and M-shell 
spectral range. Comparing with a hybrid LTE opacity code 
which combines the statistical super-transition-array and fine-structure 
methods [High Energy Density Phys., 7 (2011) 234], 
impressively good agreement is found between the two calculations.
In addition, the sensitivity of the opacity results in various plasma temperatures 
and mass densities is discussed. The ionized population fraction 
and average ionization of the Ge plasma are also described. 
Comparisons of STA results in the observed spectrum 
and opacity are considerably close while offering the advantage 
of computational speed and its capability of treating hot and dense 
high-Z plasmas.
\end{abstract}

\maketitle

\section{Introduction}
In most indirect-drive inertial confinement fusion (ICF) schemes, 
the production of high-quality thermal X-rays sources inside a 
hohlraum is of great interest. Production of X-rays comes from 
the absorption of the laser beams in the hohlraum. 
Soft X-rays propagating within the cavity between the 
fuel capsule and the inner walls will be absorbed by the 
low-Z capsule, rapidly ablating the capsule material. 
The irradiation also compresses the DT fuel inside the capsule. 
Compressing a target to ignition conditions is very challenging 
and is yet to be fully realized in experiments. 
For example, in addition to soft X-rays, energetic X-rays tend to
propagate ahead of the ablation-front and preheat the inner layer 
of ablator next to the fuel, introducing a sizable density gradients
at the boundary between the fuel and ablator.  
The density gradient excites the growth of hydrodynamic 
instabilities. A carefully tailored ablator with mid-Z materials
can potentially limit this instability effect. The hard X-rays 
can be absorbed due to the ablator opacity,  preventing 
preheating of the inner plastic layer next to the fuel, 
suppressing the instabilities (e.g., Richtmyer-Meshkov and 
Rayleigh-Taylor) along with reducing the mixing of
dense cold plasma with the less dense hot spot. It turns out 
that both germanium and silicon are hopeful choices as dopants 
in the ablator for the ICF targets. Knowing their optimum concentration 
is also essential as too much or too little will defeat its purpose. 
As a result, knowing the opacity and emissivity of Ge (and Si) 
is highly desirable.  

Experimental measurement of the emission spectra of germanium plasma 
has been reported by Hoarty {\it et al.} \cite{Hoarty2007,Hoarty2010,Harris2010}. 
In \cite{Hoarty2010}, a high-power, Gaussian laser-pulse has been used to create 
a high-temperature, high-density plasma to study its X-ray opacity and equation of states. 
Their target sample was composed of a 50/50 mixture (by particle-number) of Ti/Ge in a 50 $\mu$m 
diameter and 0.1 $\mu$m thick disc. The Ti/Ge experimental data were recorded using 
both time-integrated and time-resolved spectrometers, but only the time-integrated spectrum 
was absolutely calibrated. This latter spectrum recorded the Ge emission intensity 
in the 1.3$-$2.5 keV spectral range covering 2p$-$3d and 2p$-$4d transitions in charge 
states up to $+$29. The germanium plasma conditions were inferred from simulations
performed using collisional-radiative code FLYCHK \cite{FLYCHK}. 
In order to obtain the best fit between the calculation and experimental spectra, 
and to determine that the Ge plasma conditions is approximately 
800 eV $\pm$ 100 eV and 1.0 g/cm$^3$, a set of five calculations 
at a density 1.0 g/cm$^3$ and electron temperatures of 
700, 750, 800, 850, 900 eV were equally averaged 
to take into account the effect of temperature. 

A comparison between the local thermodynamic equilibrium 
(LTE) calculations and the measured Ge emission spectrum 
has also been discussed in the work of Harris {\it et al.} \cite{Harris2010}. 
Three LTE opacity codes, namely, GRASP2K \cite{GRASP2K}, 
CASSANDRA \cite{CASSANDRA} and DAVROS \cite{DAVROS},
have been used to calculate the synthetic Ge emission spectra. 
Each of these codes has different levels of sophistication in calculating 
the atomic structures, and hence the atomic properties. 
In Harris {\it et al}, all the LTE calculations have considered both 
the density and temperature effects. Their best fit between
the theory and experiment was obtained by averaging calculations 
among temperatures from 600$-$700 eV and densities from 1$-$2 g/cm$^3$. 
Despite the small difference in the emission intensity, the authors showed 
the GRASP2K computed widths and positions of the spectral features 
fit the experimental data well. The authors also 
compared the results from two other opacity codes CASSANDRA 
and DAVROS with the measurement. Reasonable agreement 
was reported, albeit the two UTA codes lack in spectral details in comparison 
to the GRASP2K code. Although CASSANDRA and DAVROS 
codes do show some differences in some photon-energy regions, 
their results are broadly similar. This difference was discussed 
and attributed to the variations of effective potential models 
used in CASSANDRA and DAVROS in calculating the total 
ionic energies. The comparison of emission results from LTE 
and NLTE calculations indicated that the LTE codes gave 
a temperature of about 20\% lower than the NLTE codes. 

Recently, we have employed the STA method 
\cite{STA1,STA2,STA3,STA4,STA5, STA6, STA7} 
in conjunction with the MIX model \cite{mix}
to examine the emissivities, opacities, and degree of 
average ionization for carbon and plastic CH 
in the warm, dense matter regime. We assessed the quality of 
our STA calculations by comparing them with other available 
theoretical calculations as well as experimental data. 
The STA calculated emissivities, opacities and average ionization 
for carbon and plastic CH were found to be in good agreement with 
other theoretical results and experimental data \cite{Lee2018}. 
The STA method was designed to analyze unresolved spectra of hot, 
high-Z plasma in LTE. 

In this work, we employ the STA method to study 
the radiative and atomic properties from high temperature 
and dense germanium plasma not only to validate the STA 
results for laser-produced germanium spectra, but to also investigate 
the temporal variation in density and temperature of the 
plasma through its emission process. Additionally, we also 
want to examine the STA computed multi-frequency opacity, ionized 
population fraction and average ionization, and compare them 
with the calculations obtained from the hybrid LTE opacity 
SCO-RCG code \cite{sco-rcg} for a wide range of germanium 
plasma conditions covering the L- and M-shell spectral range.   
The paper is organized as follows. In Sec. II, we outline the basic concepts 
of the STA method. In Sec. III, we present and discuss the 
STA emission spectra, opacity, ionized population fraction 
and average ionization results in comparison with other theoretical 
calculations and experimental data. We give some conclusions from
this study in Sec. IV. Unless otherwise stated, all quantities are expressed 
in atomic units. The solid density for germanium is $\rho_o$ = 5.323 g/cm$^{3}$. 
Lastly, we refer, interchangeably, to STA method as an STA model, 
or STA code throughout the paper.  

\section{Opacity model}
For our opacity model, we have adopted the STA method of 
Bar-Shalom {\it et al} \cite{STA1}. Since its first release, 
the original STA code has been modified, updated and made more stable. 
The changes made include new algorithms for better convergence 
of the computation of the partition functions \cite{Bus2013}, 
better interface between STA and MIX codes \cite{mix}, several options 
to improve the convergence of the balance between bound and free 
electrons charge densities, a better way to calculate the parametric potential 
and simplification of free-free scattering \cite{MK2019}, to name a few.

The followings outline the essential concepts of the STA method. A supershell, $\sigma$, 
is the union of energetically adjacent ordinary atomic subshells, $s \in \sigma$, where $s \equiv \bm j \equiv \{n_s, l_s, j_s\}$ .
A superconfiguration (SC) $\Xi$ of a $Q$ electron ion is defined by its supershell occupation numbers
$Q_{\sigma}$ and constructed through grouping the neighboring (in energy) ordinary subshells
into supershells. Symbolically, the expressions for $\Xi$ and $Q$ are
\begin{eqnarray}
\Xi \equiv \prod_\sigma \sigma^{Q_\sigma},~~~~~~~Q = \sum_\sigma Q_{\sigma}, 
\end{eqnarray}
respectively. This implies that the SC is constructed by partitioning the 
$Q_{\sigma}$ electrons occupying supershell $\sigma$
among the ordinary subshells in all possible ways according to $\sum_s q_s = Q_{\sigma}$, where
\begin{eqnarray}
\sigma^{Q_\sigma} \equiv  \sum_{\sum_s q_s = Q_\sigma} \prod_s \bb j^{q_s}_s
\end{eqnarray}
and $q_s$ are the occupation numbers of the subshells. 
Each partition of $Q$ in Eqs.(1) and (2) is an ordinary configuration $C \equiv \prod_{\sigma} \prod_{s  \in \sigma}\bb j^{q_s}_s$.
An ordinary configuration $C$ is a special case of SC in which each supershell contains only one shell. 
For example, an ordinary configuration:
$C = 1s^2 2s^2 2p^2_{1/2}2p^3_{3/2}$, supershell: $\sigma = (3s3p_{1/2}3p_{3/2}3d_{3/2})$ and 
a superconfiguration:
\begin{eqnarray}
\Xi = (1s)^2(2s2p_{1/2}2p_{3/2})^7(3s3p_{1/2}3p_{3/2}3d_{3/2})^1
\end{eqnarray}
is made of three supershells associated respectively with 2, 7 and 1 electrons. Note that a reasonable number of SCs 
(typically on the order of a few hundreds for mid-Z elements) can already contain a tremendous number 
of ordinary configurations. Precision for photoabsorption spectra can be improved by subdividing these SCs. 
An array $A_{CC'}$ connecting two configurations $C-C'$ can be identified 
by specifying the initial configuration $C$ and the electron(s) jumps which lead to $C'$. 
Similarly, the array $A_{\Xi \Xi'}$ connecting two SCs can be identified by specifying 
the initial configuration $\Xi$ and the electron(s) jumps which lead to $\Xi'$. 

To evaluate the STA moments 
(i.e., the total intensity, the average energy, and variance) 
and SC average rates, one needs expressions for
the populations of the configurations and super configurations.
Assuming local thermodynamic equilibrium, all the configurations 
described by a SC $\Xi$, the population of any array of levels $i$ can 
be expressed through the Saha-Boltzmann's law, $U_Q/U = N_Q/N$, 
where $N_Q$ and $N$ are the partial and total ionic number density, respectively, 
and $U_Q$ and $U$ are their corresponding partition functions. 
For example, for an ion with $Q$ electrons, the partition function 
of the SC $\Xi$ can be expressed in terms of a summation over all
levels $i$ of all configurations $C$, i.e., 
\begin{eqnarray}
U_\Xi = \sum_{C \in \Xi}  \sum_{i \in C} g_i e^{-(E^{(0)}_i+\delta E^{(1)}_{\Xi}-Q\mu)/kT}, 
\end{eqnarray}
where the sum of configuration statistical weight is given by the sum of
product of binomials
\begin{eqnarray}
\sum_{i \in C} g_i =  \sum_p \prod_{s \in C} \binom{g_s}{q_s}, 
\end{eqnarray}
each partition $p$ is a
set of $q_s$ generating an ordinary configuration $C$ and $g_s = 2 j + 1$ 
is the statistical weights corresponding to shell $s$. We also have the relations
 $\sum_{s\in\sigma} q_s = Q_{\sigma}$ and $E^{(0)}_i = \sum_{s\in\sigma} q_s \epsilon_s$, 
 where the latter quantity is the zeroth order energy. The SC energies can be written as
\begin{eqnarray}
E_{\Xi} = \delta E^{(1)}_{\Xi} + \sum_{\sigma} \sum_p \sum_{s \in \sigma}  q_s \epsilon_s, 
\end{eqnarray} 
where $\delta E^{(1)}_{\Xi}$ is the SC first-order average energy 
 correction (see Eq.(86) in Ref.\cite{STA5}), $\epsilon_ s$ the monoelectronic energy 
 and $ \sum_p$ means a summation over all 
the terms of the partition function of $Q_{\sigma}$ 
(i.e. the number of ways to distribute $Q_{\sigma}$ 
electrons in the different subshells of $\sigma$).
It should be noted that it has been demonstrated in Ref.\cite{STA1, STA5},
that with a modified set of statistical weights and supershell occupation numbers, 
the STA moments and the non-LTE average transition rates can be expressed 
in terms of generalized partition functions. 

The STA method or code was based on an ion sphere model in a chemical picture of Liberman's model \cite{inferno} 
by considering the plasma as consisting of multi-charged, multi-ionized atoms with free electrons 
shared among all ions. For a given atom with a set of the temperature and the density 
of interest, the code first solves a finite-temperature Thomas-Fermi-Dirac equation \cite{TFD1, TFD2} 
and using the solutions in terms of relativistic wave functions provides the average ionization charge state $\rm\bar Z$ 
self-consistently with the free electrons in the ion sphere. A parametric potential \cite{PP} has been used to describe 
the bound electrons as it simplifies and yet captures the changes of the electronic potential for each ion stage \cite{KM77}. 
The STA code, in the first iteration, starts by loosely defining very broad supershells, 
similar to that of the average atom model approach \cite{STA1}. The SCs are then constructed 
with these supershells. The moments of the STA transitions are then computed and the resulting 
spectra are obtained by adding up all the STA contributions. Then, in the next iteration, the supershells 
are split to optimize the corresponding SCs and this procedure repeats itself until the converged spectra 
are achieved. The potential for each SC is also progressively refined, the STA recomputed, 
and finally, the UTA moments \cite{UTA} are incorporated in the spectral-opacity calculation, 
as part of the obtaining accurate STAs' widths and energies. 

\section{Results and Discussion}
In this study, the FastRad3D code is used in conjunction with the STA opacity code 
to model the experiments of Hoarty {\it et al} \cite{Hoarty2010}. In this case, it is used 
to calculate the time-dependent temperature and density profiles of the Ge plasmas 
required by the STA as inputs. FastRad3D is a three-dimensional, 
Eulerian-based radiation-hydrodynamics code that is used primarily to simulate 
laser-matter interactions in the context of inertial-confinement-fusion experiments. 
The code couples models of hydrodynamics, laser deposition, thermal conductivity and various 
equations of state to a multi-group diffusion model for radiation transport 
and has been tested extensively against various theoretical benchmarks 
and laboratory results.  Additional details about the FastRad3D code 
can be found in Ref.\cite{fast3d}.

In the work of Hoarty {\it et al} \cite{Hoarty2010}, 
the emission spectrum is from a 0.1-$\mu$m Ti/Ge disc 
irradiated by a high-intensity (i.e., 10$^{17}$$-$10$^{19}$ W/cm$^2$), 
0.5 psec FWHM Gaussian laser pulse. Under such conditions, we hypothesize that
the laser energy is deposited in the target in a small volume 
and over a very short period of time ($\sim$ a few psec) 
so that the heating is nearly isochoric. Since the total energy 
deposited depends upon the net absorption of the extremely 
high-intensity laser pulse $-$ which was not specified 
in Hoarty {\it et al} for the displayed Ge spectrum $-$ 
we performed a series of simulations that began 
with different absorbed energies. If all of the energy was absorbed in 
$1/2$ psec at 1.0 $\times$ 10$^{18}$ W/cm$^2$, 
this would correspond to 500 kJ/cm$^2$ absorbed in the target. 
We found correspondence with the temperatures reported 
by Hoarty {\it et al} (e.g., 0.5$-$1.0 keV) for absorbed energies about 
an order of magnitude less (e.g., corresponding 
to absorbed intensities of $\sim$ 10$^{17}$ W/cm$^2$).

The absorbed laser energy is deposited uniformly 
(i.e., constant energy per electron) throughout 
the sandwich target. The energy was deposited 
primarily into the electrons because the initial, {\em impulsive} 
heating \cite{fih} was deemed to be due to hot electrons 
from either from direct inverse bremsstrahlung 
heating and/or supra-thermal electrons 
due to laser-plasma instabilities. As the electron-ion 
thermal equilibration time near-solid density 
is very short ($\sim$ psec), the target ions quickly 
equilibrate to the electrons. The evolution of the target
consisted primarily of anisotropic expansion modified 
somewhat by the radiative cooling (for example, 
our simulations show that in the first 10 psec after 
heating the target loses $\sim$ 25$\%$ of its 
initial energy to radiation). In other words, 
the laser deposition occurs on a time scale 
that is much shorter than the hydrodynamics 
response time and modeling such ``hyper-fast", 
initial dynamics poses a serious challenge 
for any radiation hydrodynamics codes. 
Consequently, we make an {\em ansatz} 
of instantaneous depositing the initial energy into the 
plasma electrons in FastRad3D 
in order to estimate the ensuing 
density and temperature profiles of the target. 
         
\subsection{Emission: theory versus experiment}
The intensity $I(\nu,T_e)$ of an emission spectrum, 
defining in units of W/(keVcm$^2$srad), can be expressed as  
\begin{eqnarray}
I(\nu,T_e) = S(\nu,T_e) (1-e^{-\rho \kappa_{\nu} L}),
\end{eqnarray}  
where $S(\nu,T_e)$ is known as the source function, that is,
 the ratio of the emissivity $\epsilon_{\nu}$ 
to opacity $\kappa_{\nu}$, $\nu$ is the frequency of the photon, 
$\rho$ is the mass density of the target matter, $T_e$ is the temperature of the plasma
and $L$ size of the target. 

\begin{figure}[!htp]
\centering
%\begin{subfigure}
\subfloat[][]
{\includegraphics[width=10cm, height=7.5cm]{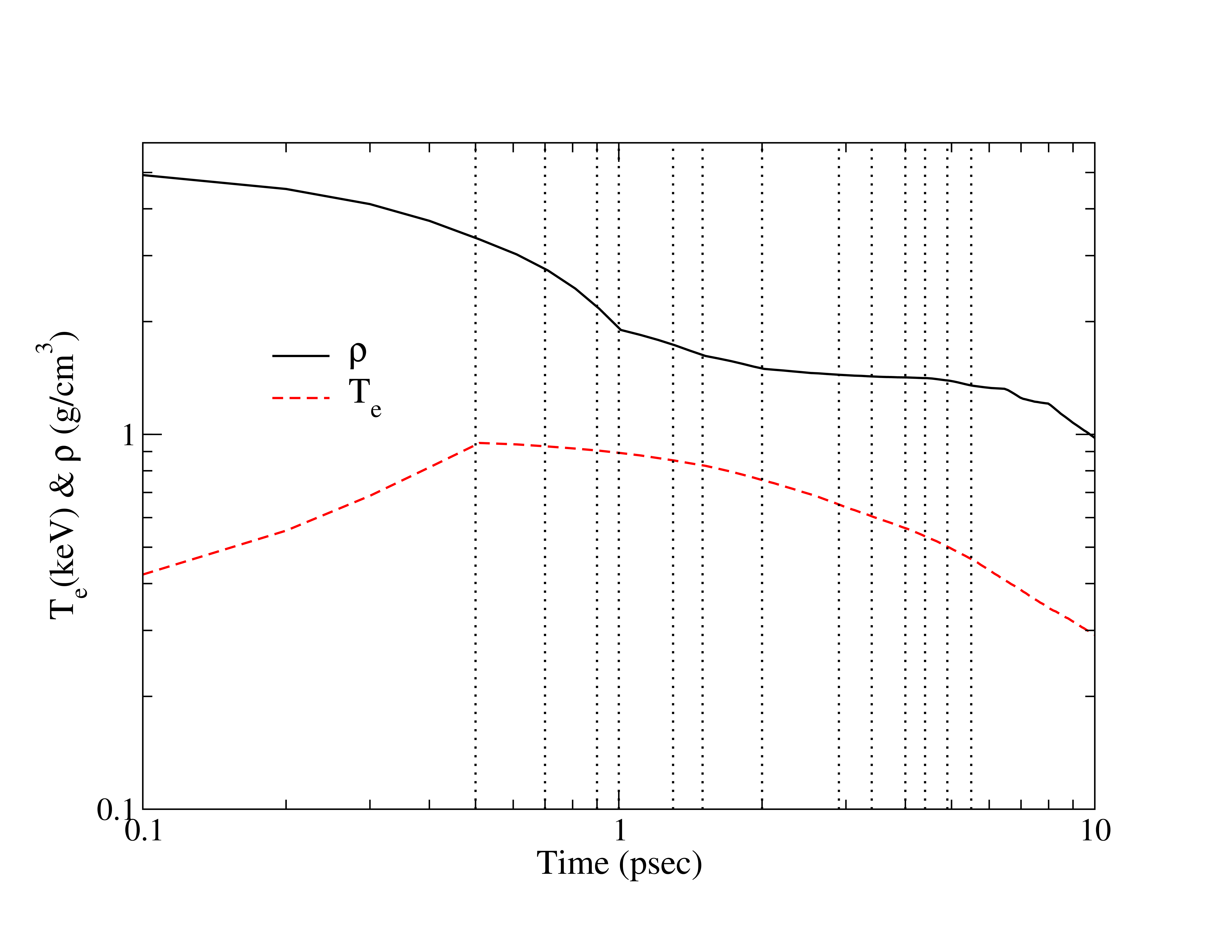}} \\
%\end{subfigure}
\vspace{0.0in}
%\begin{subfigure}
\subfloat[][]
{\includegraphics[width=9cm, height=7cm]{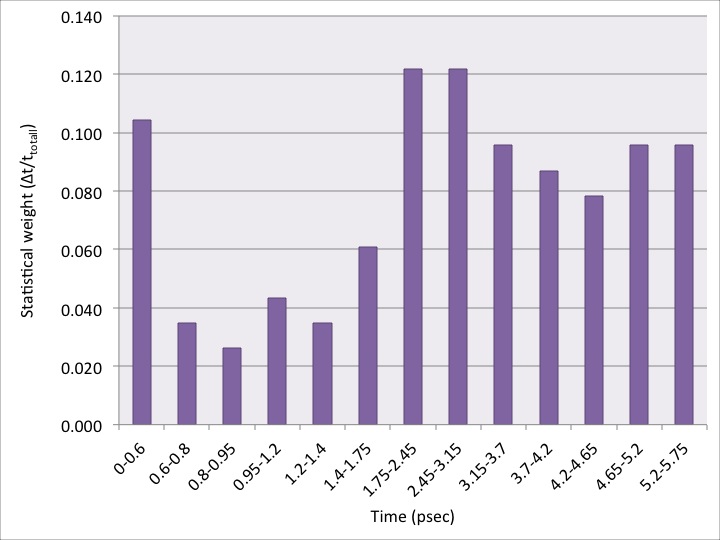}}
%\end{subfigure}
\caption{(Color online) (a) FastRad3D computed 
plasma temperature and density profiles as a function of plasma-evolution time. 
The vertical-dotted-line indicates the time snapshot, which is given in the legend of Fig.\ref{fig2}(b). 
(b) Statistical weight ($\Delta t/t_{total}$) at each $\Delta t$ corresponding to a particular plasma temperature and density. 
We choose $t_{total}$ = 5.75 psec. The sum of all the statistical 
weights amounts to 1.0.}
\label{fig1}
\end{figure}

\begin{figure}[!htp]
\centering
\includegraphics[width=9cm, height=13cm]{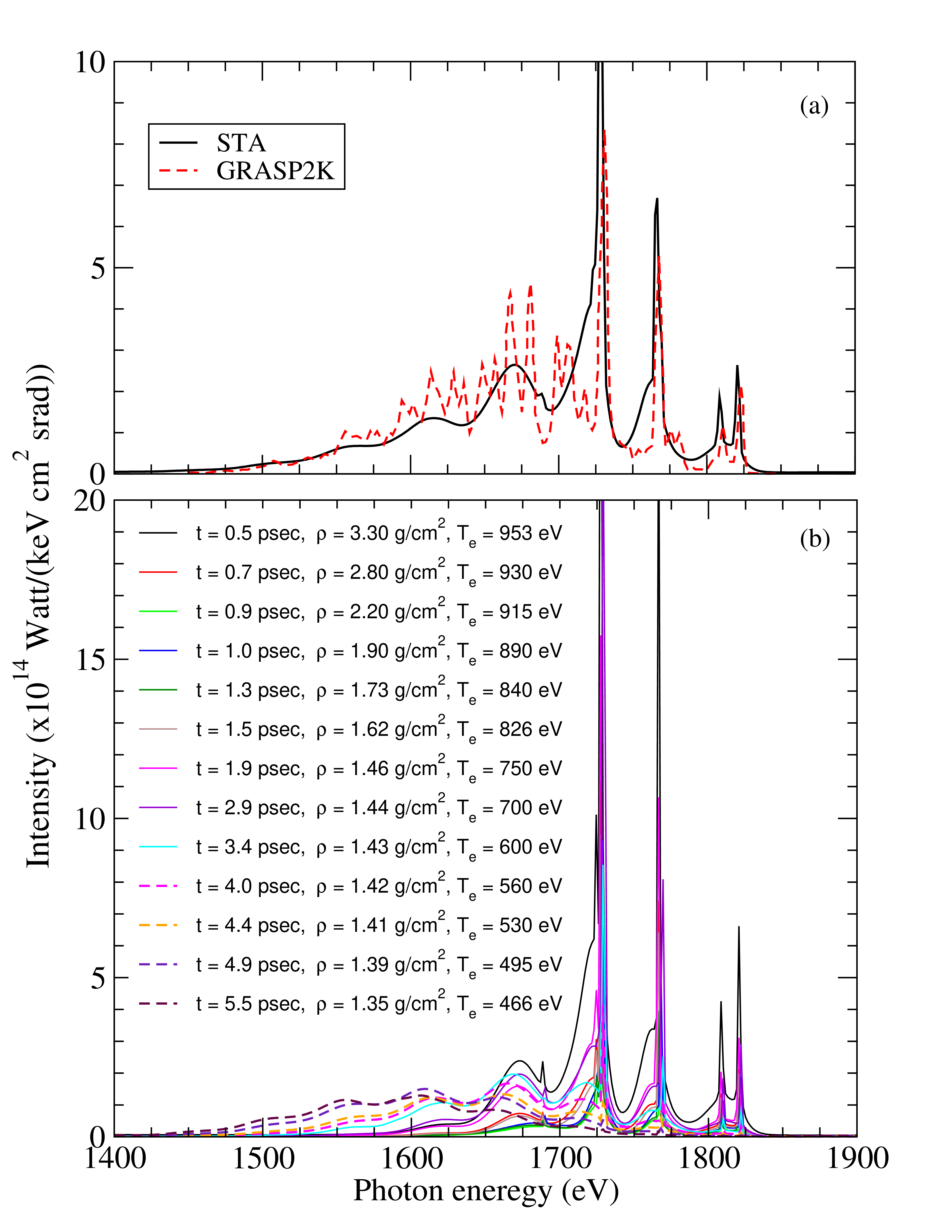}
\caption{(Color online) (a) Comparison of simulated emission from GRASP2K and STA codes.
The LTE GRASP2K calculation, digitized from Ref.\cite{Harris2010}, includes contributions 
from plasma between 600$-$700 eV and 1.0$-$2.0 g/cm$^3$ to model the affect of both 
temperature and density effects. (b) The ``time-resolved" STA emission spectra that contribute 
to the total or ``time-integrated" emission spectrum.}
\label{fig2}
\end{figure}

\begin{figure}[!htp]
\centering
\includegraphics[width=9cm, height=12cm]{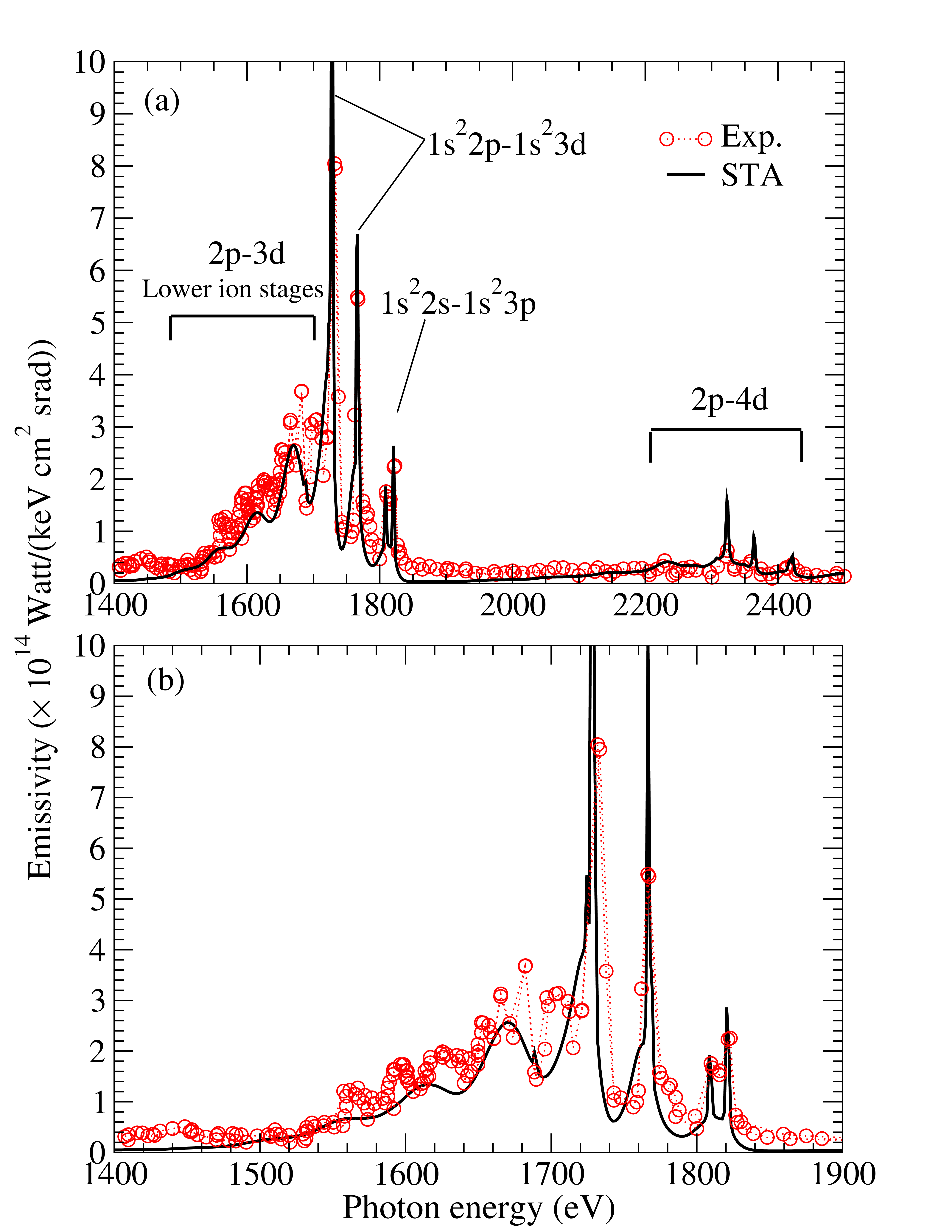}
\caption{
(Color online) Emission spectrum of germanium. 
(a) STA calculation versus the experimental data 
digitized from Ref.\cite{Hoarty2010}. 
(b) Same plot as (a) but for photon energy between 1.4 and 1.9 keV. 
Note that the 3-eV instrument-broadening is not considered it appeared   
to be insignificant to affect the main features of our spectrum.}
\label{fig3}
\end{figure}

In order to construct the time-integrated STA emission spectrum $I(\nu,T_e)$ shown later in 
Fig.~\ref{fig2}(a) and Fig.~\ref{fig3}, we have performed several 
FastRad3D calculations varying the laser-absorbed intensity of $\sim$ 10$^{17}$ W/cm$^2$.   
We find that the best fit for our case is at 0.8 $\times$ 10$^{16}$ W/cm$^2$. 
The corresponding spectra are shown in Fig.~\ref{fig2}(b). Each 
spectrum corresponds to the plasma in a particular conditions 
(i.e., density and temperature) at a particular time. The 
time evolution profiles of the plasma density and temperature
are presented in Fig.~\ref{fig1}(a).

Deducing from this time-evolution information of Fig.~\ref{fig1}(a), 
we figure out how much time the plasma spends 
at a particular density and temperature to help us to construct 
the relevant histogram shown in Fig.~\ref{fig1}(b). 
Consequently, we arrive at an expression for 
the time-integrated spectral intensity ($I_f$) as a sum of contributions 
from multiple time-weighted intensities ($I_i$) over 
the ``cooling'' period $t_{total}$: 
\begin{eqnarray}
I_f = \sum_i^N(\Delta t_i/t_{total}) \times I_i,
\end{eqnarray}
with the maximum number of time-snapshots in this case is $N$ = 13. 
The histogram basically gives the statistical weights 
(or probability) of the plasma-condition 
over the plasma cooling period. The statistical weight $(\Delta t_i/t_{total}) $ 
can be obtained by first determine the $\Delta t$. For example, 
$\Delta t_1$= $(0.5\times(0.7-0.5)+0.5)-0.0)$ psec = 0.6  psec,  
$\Delta t_2$ = $(0.5\times(0.9-0.7)+0.7)-0.6)$ psec = 0.2 psec, 
and so forth, and hence their corresponding statistical weights 
are simply $\Delta t_1/t_{total}$ = 0.6/5.75 = 0.104, 
$\Delta t_2/t_{total}$ = 0.2/5.75 = 0.035, and so forth. 
Note that we chose the cut-off time to be at $t_{total}$ = 5.75 psec since 
we found the spectral contributions from later times, 
namely, $t$ = 6.0 and 7.0 psec, to be relatively insignificant. 

It is informative to compare STA results with those from other LTE opacity codes. 
In Fig.~\ref{fig2}(a), we have the STA computed spectrum superimposed onto the 
result from LTE opacity code GRASP2K \cite{GRASP2K, Harris2010,Hoarty2010}. 
GRASP2K is a general-purpose relativistic atomic structure package based on a fully 
relativistic multiconfiguration Dirac-Hartree-Fock method. 
It performs detailed term accounting (DTA) calculations to produce energy levels 
and oscillator strengths which can then be used to construct a spectrum using 
Saha-Boltzmann statistics. The code is well-known for its accuracy in 
determining the line positions and widths, and has been considered by many atomic 
physicists as a standard for benchmarking. However, if GRASP2K 
is similar to the HULLAC \cite{hullac} in the sense that it is an ``isolated-atom" code,
which does not take into account the density effects by allowing 
for the ionization potential depression, its results in this case
can be less accurate. 

Fig.~\ref{fig2}(a) also shows that the STA calculated spectrum 
matches the locations satisfactorily, including the peaks for the $2p-3d$ 
and $2p-3p$ transitions obtained by GRASP2K but provides 
envelopes to the detailed results for the lower ion stages.  
This is because the STA approach sometimes cannot resolve 
spectral features that are simply not made up of single transition, 
particularly in the case for lower ion stages of $2p-3d$ transitions, 
but are coalescences of similar transitions from several ionization stages. 
In fact, the number of transitions in a feature can be so large that statistical 
treatments provide a method of determining the spectral feature characteristics. 
For example, see the discussions of unresolved transition arrays (UTA) by 
Bauche, Bauche-Arnoult and Klapisch \cite{UTA}.
Under some circumstances, such a statistical approach may overestimate the 
Rosseland mean opacity because all the gaps between the lines can be overlooked. 
In any case, it is important to note that the spectrum from the GRASP2K 
simulations matched the experimental data well and placed the plasma conditions at 
1.5 g/cm$^3$ $\pm$ 0.5 g/cm$^3$ and 600 eV $\pm$ 60 eV. On the other hand,
result from the non-LTE collisional-radiative equilibrium code FLYCHK 
also shows a good match to the experimental spectra but indicates 
$T_e$ $\approx$  800 $\pm$ 100 eV and $\rho$ $\approx$ 
1.5 $\pm$ 0.5 g/cm$^3$. Given that the results of collisional-radiative   
FLYCHK calculations also got good agreement with the experimental 
data lead us thinking that the plasma could lie in this part-LTE, part-non-LTE domains. 
Of course, additional emission spectra from both CASSANDRA \cite{CASSANDRA} 
and DAVROS \cite{DAVROS} opacity codes for further comparison would also be 
beneficial. However, we are unable to extract the clean digitized data of CASSANDRA 
and DAVROS calculations from the published figures \cite{Harris2010}.  

The time-snapshots of different contributions from
the final spectra shown in Fig.~\ref{fig2}(b) illustrate that the strong 
emission profile appears between 1.7 and 1.8 keV comes 
from the initial time $t <$ 2.0 psec contributions. 
On the lower temperature side, the notable emission appears 
in the 1.5$-$1.7 keV region comes from the later time 3.0 $< t <$ 5.0 psec 
contributions.

Figure \ref{fig3} displays a comparison between the synthetic Ge spectrum 
from STA calculation and the measured spectrum in the photon-energy range 
of 1.3$-$2.5 keV. In this photon-energy range, the spectrum reveals the L-band 
transitions. The STA spectrum appeared to be in reasonably good agreement with the 
experimental data. In particular, the STA reproduces the dominant features of $2p-3d$, 
$1s^2 2s-1s^2 3p$ and $1s^2 2p-1s^2 3d$ and $2p-4d$ transitions depicted 
in the measured Ge emission spectrum. However, magnifying 
the spectra in the photon energy-range between 1.4 and 1.9 keV for detailed 
comparison in Fig.\ref{fig3}(b) apparently reveals some disparities
between the synthetic and experimental spectra. This could mean
the presence of spatial temperature and density variations in the sample 
as well as the non-LTE effects. The limitation of STA to account for these 
effects and to quantitatively modeling the experimental data is expected 
and underscoring the difficulty of the present attempts. Besides, the model 
assumed local thermodynamics equilibrium population dynamics. 
From this standpoint, the good agreement between the 
GRASP2K results and experimental data reported in Ref.\cite{Harris2010} 
(e.g., see Figs. 2 and 3) is puzzling and most likely be fortuitous as GRASP2K 
itself is a LTE model. Furthermore, recent work has also shown these to have 
a greater effect than was previously thought \cite{Hoarty2017}. In any case, 
if we were asked to estimate the ``time-integrated" plasma conditions, 
according to our calculations we say our estimates to be around 
$\rho~\approx$ 1.7 g/cm$^3$ and $T_e~\approx$ 690 eV which 
are quite close the values reported in Ref.\cite{Harris2010}.

\begin{figure}[!htp]
\centering
\includegraphics[width=10cm, height=8cm]{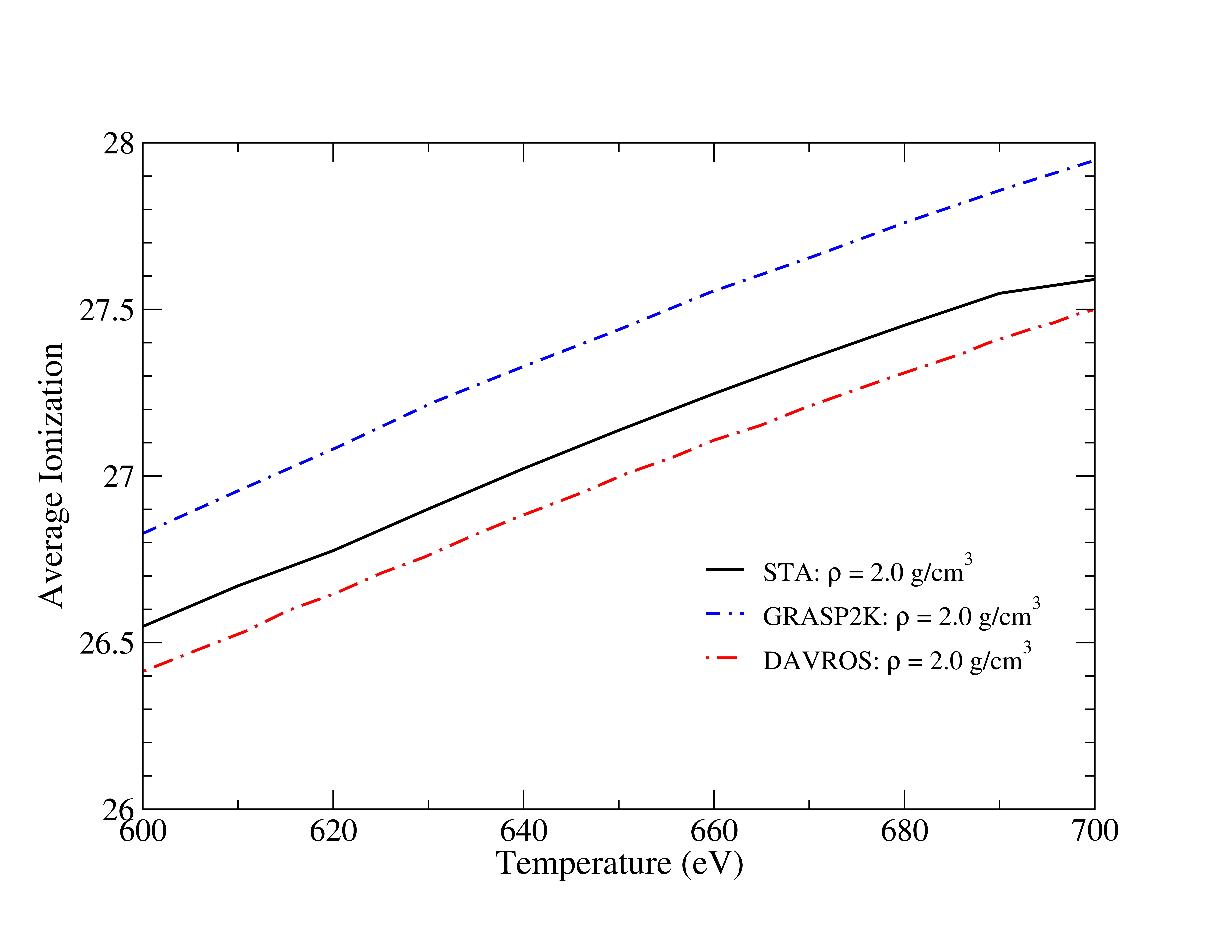}
\caption{(Color online) Average ionization versus plasma temperature. 
Note that results from GRASP2K and DAVROS are digitized from Ref.\cite{Harris2010}.}
\label{fig4}
\end{figure}

In Fig.~\ref{fig4}, we further compare the STA, GRASP2K, and DAVROS calculated average 
ionization versus the electron temperature at a mass density of 2.0 g/cm$^3$. 
Notice that, in this temperature range, more than two-thirds of Ge are ionized. 
Raising the temperature from 600 to 700 eV increases the charge 
state of Ge ions by almost two units. By the way, we have left out the 
average ionization result from CASSANDRA calculation at 2.0 g/cm$^3$ 
because the CASSANDRA result is essentially identical to those of DAVROS. 
The plot also shows that the STA's average ionization values are
very close to that of DAVROS $-$ the difference between 
the two curves is less than 1\%. Alternately, comparing the 
average ionization values between the STA and GRASP2K models, 
it is obvious that GRASP2K simulation shows the largest deviation 
from STA or DAVROS. This difference has 
been shown and discussed in \cite{Harris2010} that it was thought that a relatively 
small number of configurations was used in the Saha-Boltzmann 
partition function of GRASP2K in determining the average ionization values. Now, concerning 
the difference observed between the STA and DAVROS. Possible causes may be due to
(i) the statistical approximation used in the STA model to amalgamate the similar transitions 
from several ionization stages and (ii) the use of parametric potential \cite{PP} 
in place of the Hartree-Fock potential in our numerical solution of the Dirac's equation. 

\begin{figure}[!ht]
\centering 
  \subfloat[$\rho \le$ 1.0 g/cm$^3$]{\includegraphics[width=7.5cm, height=12cm]{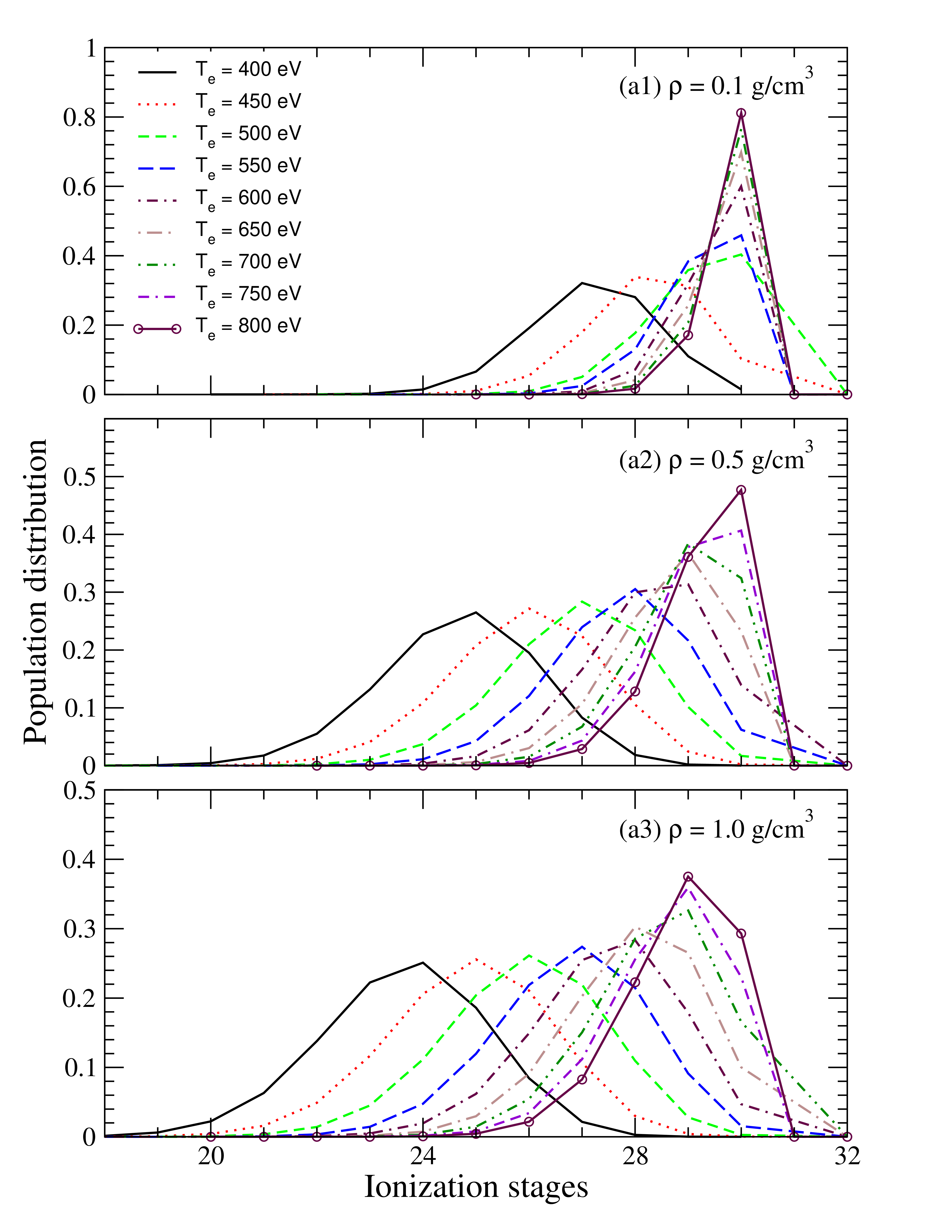}} 
  \subfloat[$\rho >$ 1.0 g/cm$^3$]{\includegraphics[width=7.5cm, height=12cm]{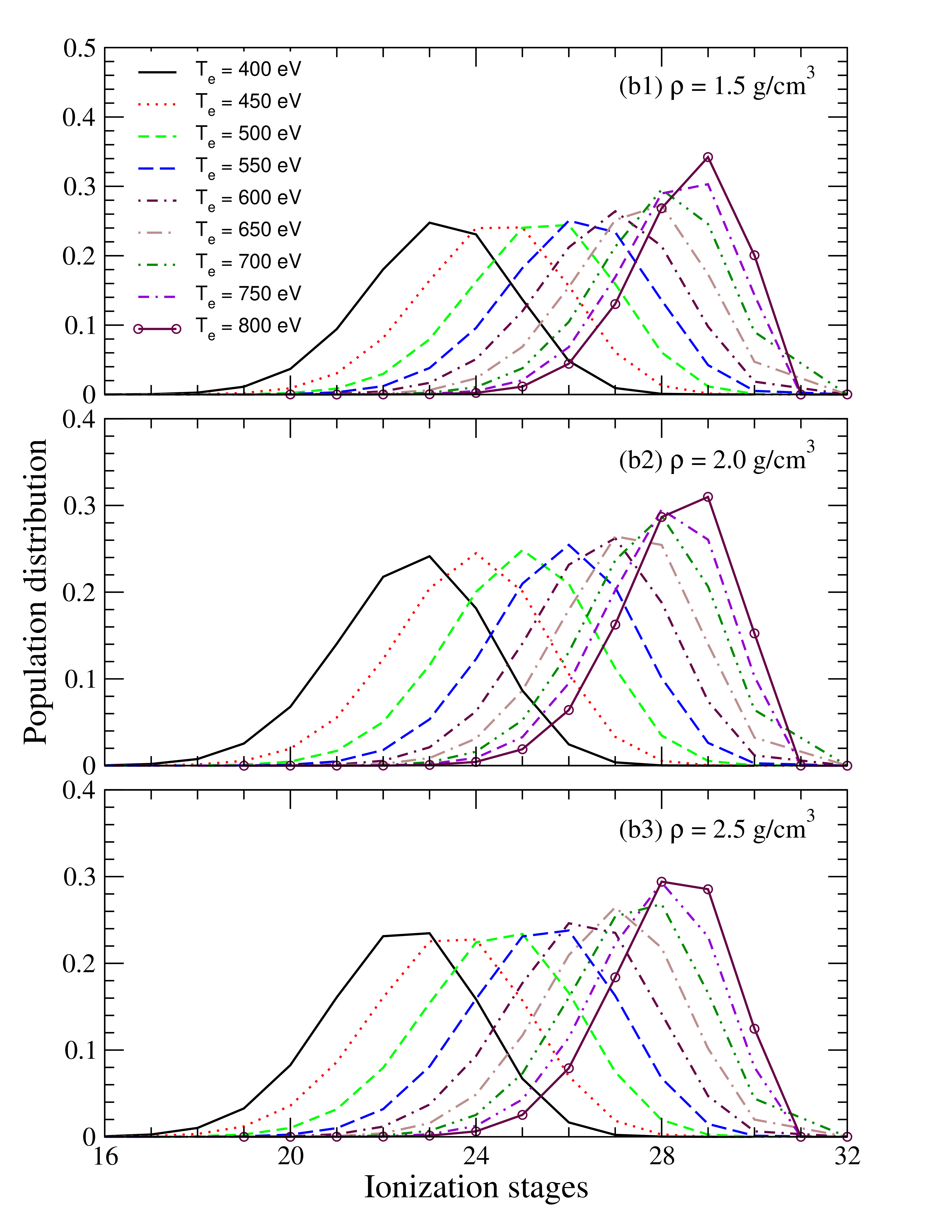}} 
  \caption{(Color online). Ionization fractions versus the charge states for Ge at various temperatures and densities.} 
\label{fig5}  
\end{figure}

Next, we examine in Fig. \ref{fig5} the LTE ionization fractions versus
germanium charge states at various temperatures and densities. Here
we see that within these temperature and density range, indeed, 
several ionization stages do contribute to the transitions we saw 
earlier and are keeping with the explanation we offered earlier 
for the discrepancy. For the case of $\rho >$ 1.0 g/cm$^3$, 
at a temperature of 800 eV the ionization balance appears 
to be near charge state of 28 with a population fraction of $\sim$ 0.32. 
Similarly, at a temperature of 400 eV the ionization 
balance moves to a lower charge state and appears 
to be near charge state of 23 with a population fraction 
drops to $\sim$ 0.25. For the case of $\rho \le$ 1.0 g/cm$^3$, 
we can see that the ionization balance strongly depends on the temperature 
and mass density. For example, at higher electron temperature of 800 eV 
the ionization balance peak-value drops by a factor of 2 and extends 
over a broader charge states as one moves from 0.1 to 1.0 g/cm$^3$.
On the other hand, at a lower electron temperature of 400 eV the ionization 
balance peak-value stays roughly the same, but its peak moves from charge 
state 27 to 24 as one compresses the plasmas (see Fig. \ref{fig5}(a1)-(a3)).  

\subsection{Opacity: theoretical models comparison}

We now turn to examine density- and temperature-dependent opacity. 
The Rosseland mean opacity, $\kappa_R$, and Planckian mean opacity, $\kappa_P$, are defined by
\begin{eqnarray}
\frac{1}{\kappa_R} &=& \frac{15}{4\pi^4}\int_0^\infty \frac{x^4e^{x}dx}{u(x)(e^{x}-1)^2}, \\
\kappa_P &=& \frac{15}{\pi^4}\int_0^\infty  \frac{u(x)x^3 dx}{(e^{x}-1)},
\end{eqnarray}
respectively, where $x = h\nu/kT_e$, $u(x) = N_A{\bar \sigma(h\nu)}/A$, $N_A$ 
is the Avogadro's constant, $A$ denotes the atomic mass number, 
${\bar \sigma(h\nu)} = \sigma(h\nu)(1-e^{-x})$ and $ \sigma(h\nu)$ is 
the total cross-section including all processes like scattering of photon, 
bound-bound, bound-free and free-free absorptions.

%%%%%%%%%%%%%%%%%%%%%%%%%%%%%%%%%%%%%%%%%%%%%%%%%%%%%%%%
\begin{figure}
\centering
\includegraphics[width=10cm, height=14.5cm]{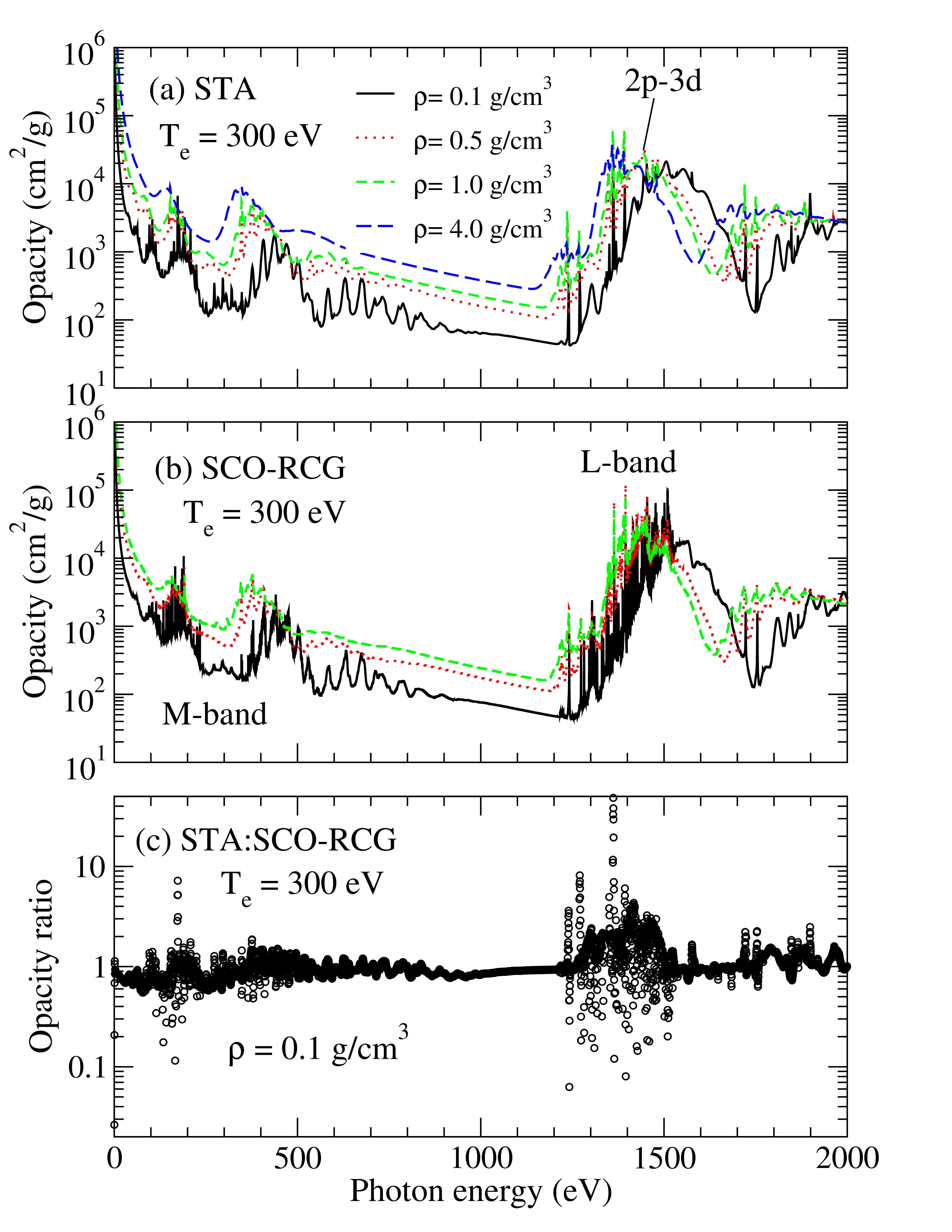}
\caption{The density dependence of the radiative opacity for germanium at electron temperature of 
300 eV are shown in (a) STA and (b) SCO-RCG. Opacity ratio of STA to SCO-RCG 
for the same temperature at $\rho$ = 0.1 g/cm$^3$ is shown in (c). The deviation from unity in opacity ratio 
is likely due to (i) the difference in photon-energy resolution used in the STA and SCO-RCG codes and 
(ii) the replacement of the superconfigurations by the detailed line-by-line treatment in the SCO-RCG code.}
\label{fig6}
\end{figure}

%%%%%%%%%%%%%%%%%%%%%%%%%%%%%%%%%%%%%%%%%%%%%%%%%%%%%%%

\begin{figure}[!htp]
\centering
\includegraphics[width=10cm, height=14.5cm]{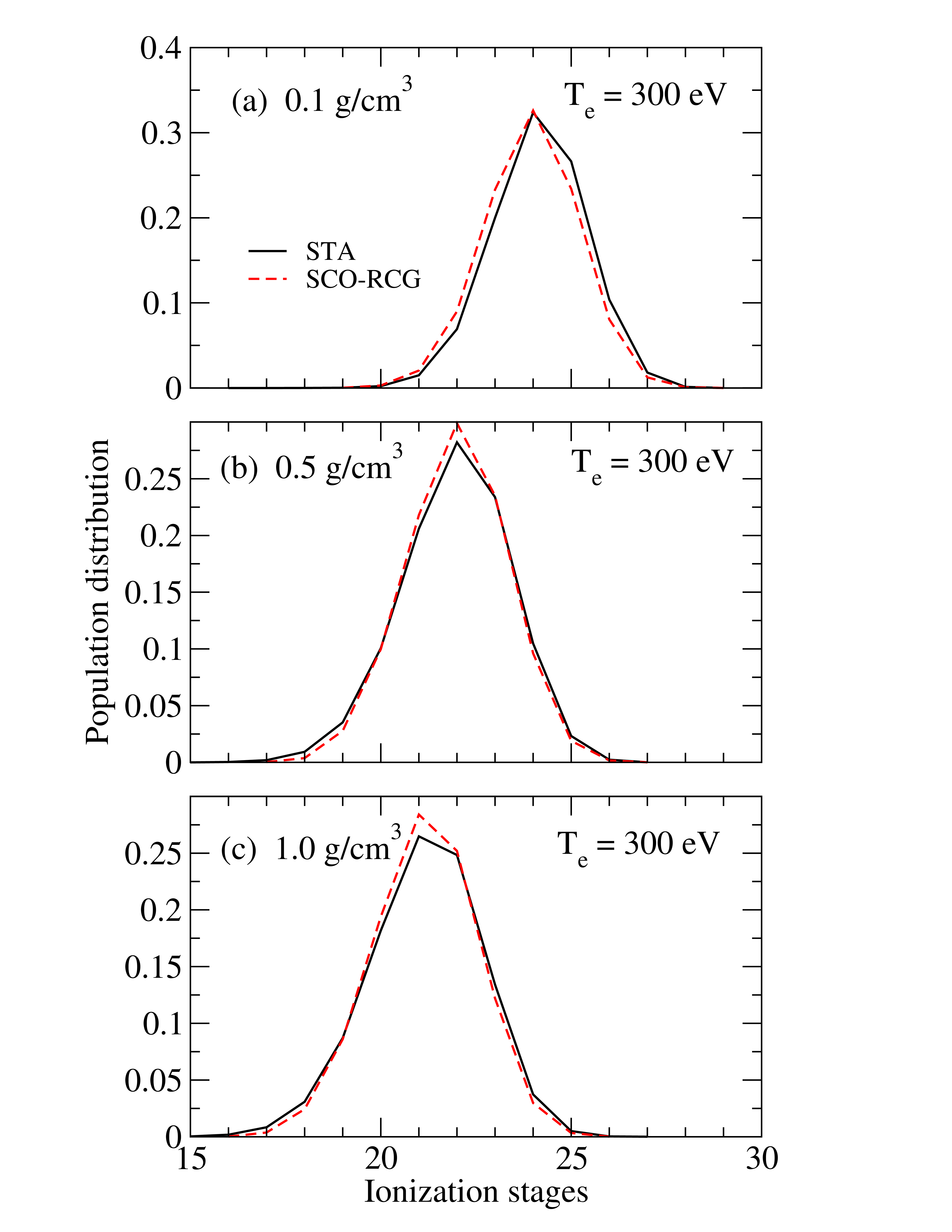}
\caption{(Color online) Ionization fractions versus the charge states for germanium at various densities.}
\label{fig7}
\end{figure}

Figure \ref{fig6} presents the opacity results of STA calculations 
together with the SCO-RCG calculations at $T_e$ = 300 eV, at various mass densities.
The plots show a good agreement between the two theoretical calculations. 
In brief, the SCO-RCG is a LTE hybrid opacity code which combines the statistical super-transition-array 
approach and fine-structure calculations for intense and spectrally broad transition arrays \cite{sco-rcg}. 
Criteria are used to select transition arrays which are removed from the super-configuration statistics, 
and replaced by a detailed line-by-line treatment. The data required for the calculation of the detailed 
transition arrays, like Slater, spin-orbit and dipolar integrals, are obtained from the super-configuration code 
SCO (Super-Configuration Opacity) \cite{SCO0}, to provide a consistent description 
of the plasma screening effects in the wave functions. Then, the level-energies, 
line-positions, and line-strengths are calculated using the RCG  routine of 
Cowan's atomic structure code \cite{cowan}.

\begin{figure}
\centering
\includegraphics[width=10cm, height=14.5cm]{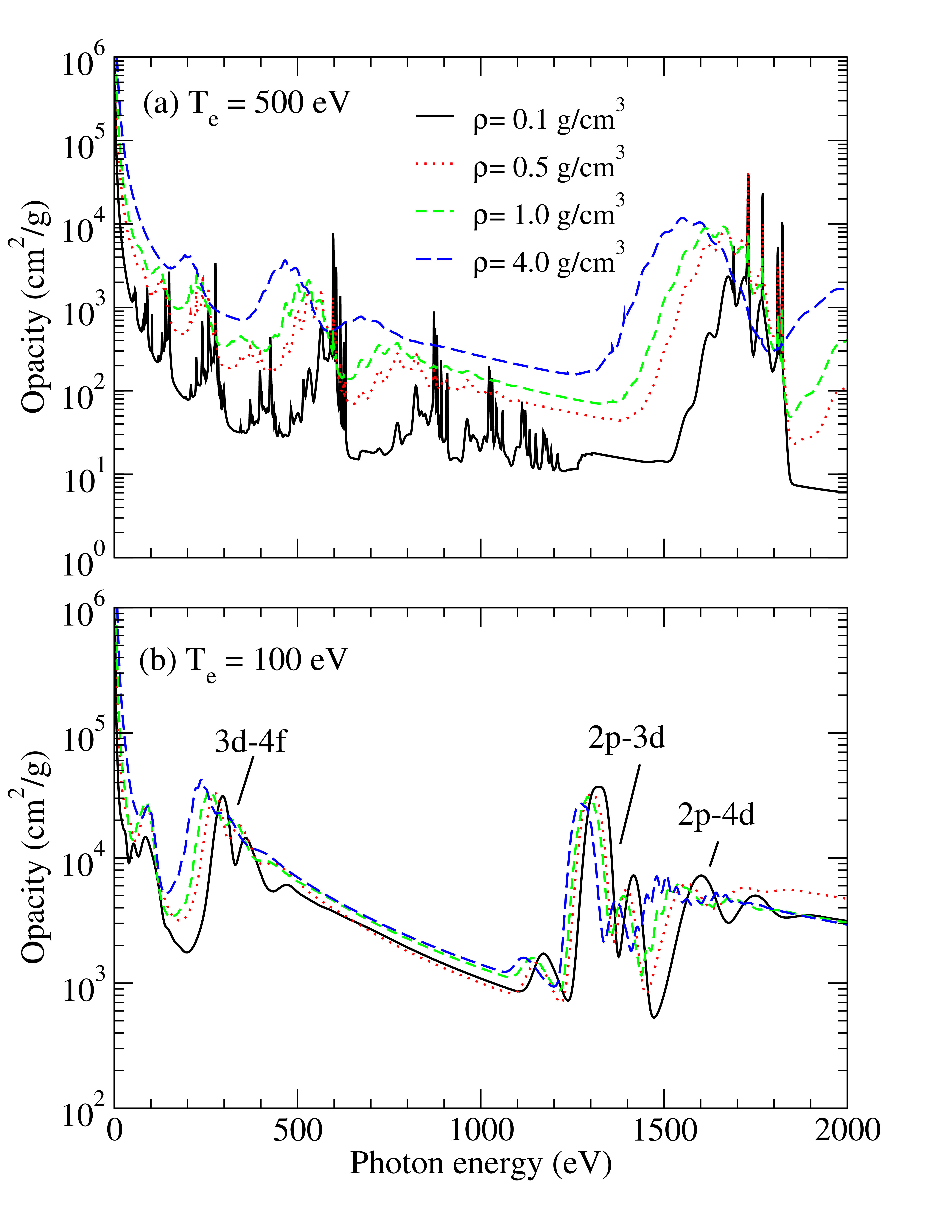}
\caption{The density dependence of the STA radiative opacities for germanium at 
electron temperature (a) $T_e$ = 500 and (b) 100 eV.}
\label{fig8}
\end{figure}

Figure \ref{fig6} also reveals two pronounced structures. In addition 
to the L-band structures with a corresponding photon energy range of 1.1 $-$ 1.6 keV, 
one also sees the M-band structures with a corresponding photon energy range of 0 $-$ 500 eV.  
The dependence of the opacity on the plasma density is obvious. 
Because of the depression of the ionization potential as the density rises, 
the Inglis-Teller limit \cite{IT} in which many spectral lines merge and show 
a quasi-edge appears to shift to lower and lower photon energies, and slowly 
disappears into the continuum. It is important to note that although the opacity 
profiles of Planck and Rosseland means of Ge plasmas 
have been investigated using the SCO-RCG opacity 
code by Benredjem {\it et al} \cite{SCO1,SCO2,SCO3,SCO4, SCO5}, this is the 
first comparative study between the SCO-RCG and STA opacity results. 
In addition to the opacity, we also compare the STA predicted ionization 
fractions at $T_e$ = 300 eV with the results from SCO-RCG calculation 
for $\rho$ = 0.1, 0.5 and 1.0 g/cm$^3$. As shown in figure \ref{fig7}, 
they are nearly identical.

\begin{figure}[!htp]
\centering
\includegraphics[width=12cm, height=8cm]{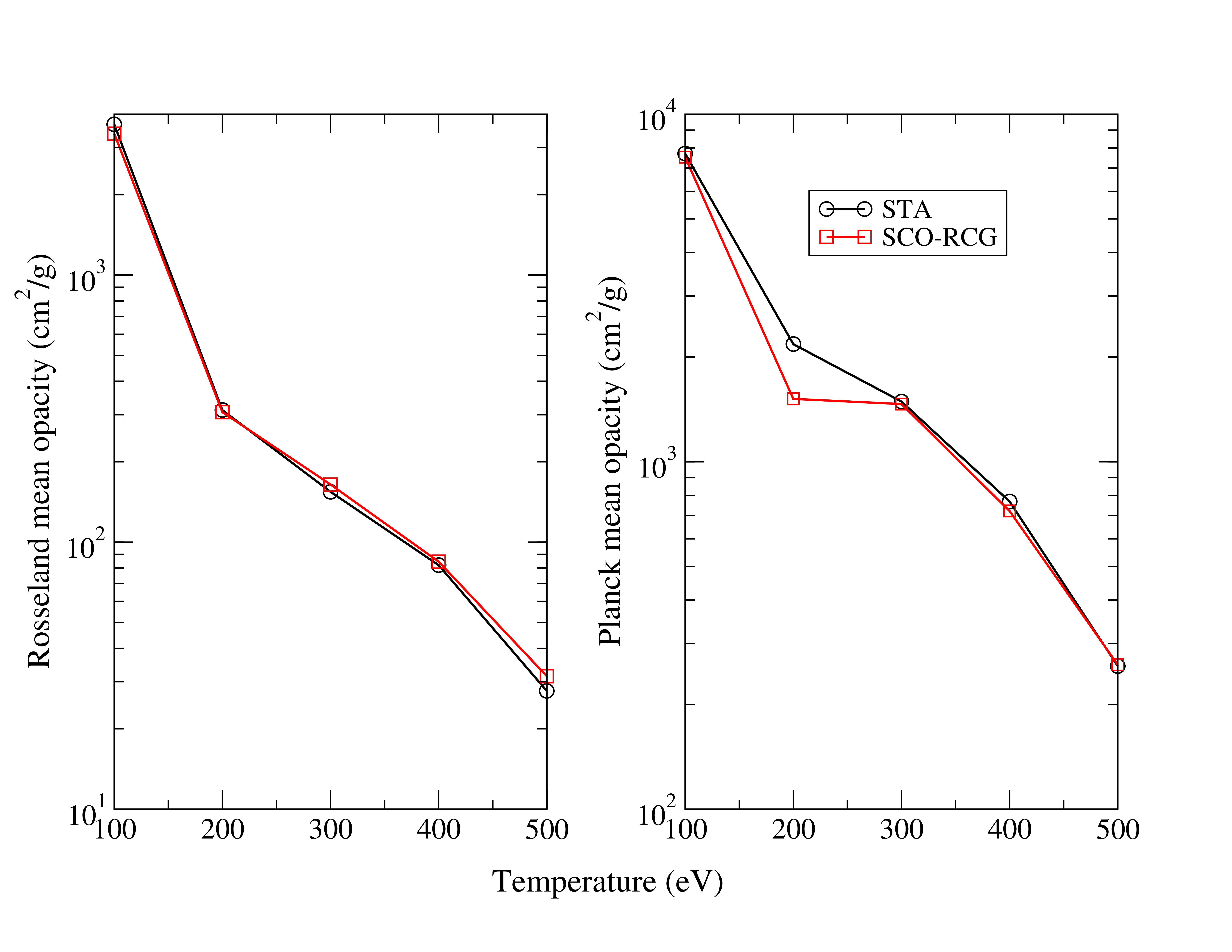}
\caption{(Color online) Mean opacity versus plasma temperature at 0.1 g/cm$^3$.}
\label{fig9}
\end{figure}

The sensitivity of the opacities to the variation of the plasma temperatures is displayed in Fig.\ref{fig8}. 
Comparing the opacities at temperatures of 500 and 100 eV, one can see that the opacity at 500 eV 
displays more complex structures than the one at 100 eV; the variation of opacity with density 
is also stronger at higher temperatures, at least for the range of density and temperature considered in this figure. 
These features have been observed and discussed recently by Mondet {\it et al} \cite{SCO1}. 
The main cause of this is due to the increase of level populations as one increases the temperature. 
One can also envisage that narrow spin-orbit separation such as between $2p_{3/2}-3d_{5/2}$ and $2p_{1/2}-3d_{3/2}$ 
will evanesce due to the thermal broadening when the temperature is increased and to an increase of level populations. 
Here we also notice that the opacity of germanium at 100 eV is higher than that at 500 eV, except around 1.7 keV.
Again, such behaviors are also consistent with the observations of Mondet {\it et al} \cite{SCO1}.

\begin{center}
%\vspace{10mm} % Adjust the height of the space between caption and tabular
\begin{tabular}{ |c|c|c|c|c|c|} 
\hline
$T_e$(eV) &~~$\rm\bar Z_{SCO-RCG}$~~&~~$\rm\bar Z_{STA}$~~
&~~$\rm\bar Z_{AA}$~~&~~$\rm\bar Z_{STA}/\rm\bar Z_{SCO-RCG}$~~
&~~$\rm\bar Z_{AA}/\rm\bar Z_{SCO-RCG}$~~ \\
\hline
100 & 15.550 & 15.837 & 15.241 & 1.0185  &  0.9801 \\
200 & 20.780 & 20.965 & 20.804 & 1.0089  &  1.0012 \\ 
300 & 23.943 & 24.141 & 23.912 &  1.0083 &   0.9987\\ 
400 & 26.903 & 27.168 & 26.912 &  1.0099  & 1.0003  \\ 
500 & 28.609 & 29.093 & 28.693 &  1.0169 & 1.0029 \\ 
\hline
\end{tabular}
\captionof{table}{Average ionization at $\rho$ = 0.1 g/cm$^3$.}
\label{tableB}
\end{center}
 
Finally, it is also of interest to compare the STA Rosseland and 
Planck mean opacities against the results from SCO-RCG \cite{SCO3}. 
As shown in Fig.~\ref{fig9}, we see that the Rosseland mean opacity results 
of SCO-RCG and STA calculations are in close agreement. As for the 
Planck mean opacity, it is shown that STA matches the SCO-RCG 
points favorably, except at the plasma temperature of T$_e$ = 200 eV 
where the STA value is about 40\% higher than the SCO-RCG one. 
The fact that the Planck value predicted by SCO-RCG at $T$=200 eV 
is significantly different from the one at $T$=100 eV, 
but rather close to the one at $T$=300 eV, can be 
explained by the populations of the different subshells of the M shell 
(see table \ref{table1}). Indeed, the transitions which 
have the strongest weights in the Planck and Rosseland means 
are 3-4, 3-5 and 3-6 transitions (see tables \ref{table3}) and 
the opacity is proportional to the populations of the initial state. 
As we can see, the variations of populations of the $n=3$ subshells 
are more important between $T$=100 and $T$=200 eV, 
than between $T$=200 and $T$=300 eV. This is particularly 
obvious for the $3s$ subshell. These population changes impact 
both the photo-excitation and the photo-ionization contributions to opacity 
(see table \ref{table4}). Of course, the 2-3 transitions also contribute 
to the mean opacities but are not responsible for the discrepancy 
between SCO-RCG and STA. Table IV shows that the photo-excitation 
is the dominant contribution to the Planck mean. Moreover, 
the difference in the STA and SCO-RCG photo-ionization cross sections 
involves low photon energies which do not affect the mean opacities. 
The difference between SCO-RCG and STA is mainly 
due to the photo-excitation (bb) contribution: 1354 cm$^2$/g 
for SCO-RCG and 2237 cm$^2$/g for STA, at $T$=200 eV 
(see Table \ref{table4}). This discrepancy might be attributed 
to the difference between the Detailed-Line-Accounting 
treatment used in SCO-RCG (which involves only ordinary 
configurations except for highly excited states \cite{Pain2015}) 
and the statistical super-transition-array model used in STA code. 
However, as mentioned above, except for that particular 
temperature, the results of STA are very close to the SCO-RCG ones, 
for a much lower numerical cost.

\begin{table}[h!]
\centering
\begin{tabular}{|c|c|c|c|c|c|}\hline
$T$ (eV) / Subshell & 2s & 2p & 3s & 3p & 3d \\\hline\hline
100 & 2.000 & 6.000 & 0.927 & 2.026 & 1.894  \\\hline
200 & 1.965 & 5.771 & 0.099 & 0.235 & 0.286  \\\hline
300 & 1.448 & 3.756 & 0.030 & 0.078 & 0.111  \\\hline
400 & 0.700 & 1.756 & 0.015 & 0.043 & 0.066  \\\hline
500 & 0.251 & 0.646 & 0.008 & 0.023 & 0.037  \\\hline
\end{tabular}
\caption{Average populations of the different subshells belonging 
to L and M shells in a germanium plasma at $\rho$=0.1 g/cm$^3$ 
and various temperatures, obtained from SCO-RCG calculations.}
\label{table1}
\end{table}

\begin{table}[h!]
\centering
\begin{tabular}{|c|c|c|c|c|c|}\hline
Transition & Average-atom energy \\\hline\hline
3d-4f & 359 eV \\\hline
3p-4d & 409 eV \\\hline
3s-4p & 434 eV \\\hline
3d-5f & 504 eV \\\hline
3p-5d & 562 eV \\\hline
3d-6f & 583 eV \\\hline
3s-5p & 600 eV \\\hline
3p-6d & 642 eV \\\hline
3s-6p & 685 eV \\\hline
2p-3s & 835 eV \\\hline
2p-3d & 1350 eV \\\hline
2s-3p & 1445 eV \\\hline
\end{tabular}
\caption{Average energies of the most important transitions involving 
L and M shells in a germanium plasma at $T$=200 eV and $\rho$=0.1 g/cm$^3$, 
obtained from SCO-RCG calculations.}
\label{table3}
\end{table}

\begin{table}[h!]
\centering
\begin{tabular}{|c|c|c|c|c|c|c|c|c|}\hline
$T$ (eV) & \multicolumn{8} {c|}{Planck mean opacity (cm$^2$/g)}\\\cline{2-9}
               & $\kappa_{\rm SCO}(\mathrm{ff})$ & $\kappa_{\rm STA}(\mathrm{ff})$ & $\kappa_{\rm SCO}(\mathrm{bb})$ & $\kappa_{\rm STA}(\mathrm{bb})$ & $\kappa_{\rm SCO}(\mathrm{bf})$ &  $\kappa_{\rm STA}(\mathrm{bf})$ & $\kappa_{\rm SCO}(\mathrm{total})$ & $\kappa_{\rm STA}(\mathrm{total})$ \\\hline\hline
100 & 97.24 & 73.47 &  6441 & 6624 &  985.6 & 1012.03 & 7523.84 & 7709.50             \\\hline
200 & 20.45 & 17.08 &1354   & 2237 & 142.7  & 122.73 & 1517.15   & 2376.81           \\\hline
300 & 7.589 & 6.477 &1389  & 1421  &  68.86  & 63.13 & 1465.45    & 1490.61            \\\hline
400 & 3.930 & 3.057 &678.1 & 730.0 &  39.41     & 35.47 & 721.44   & 768.53            \\\hline
500 & 2.162 & 1.687 & 238.1 & 241.77  & 20.29  & 14.8 & 260.55     & 258.26            \\\hline
\end{tabular}
\caption{Contributions of photo-excitation and photo-ionization to 
Planck mean opacity in a germanium plasma at $\rho$=0.1 g/cm$^3$ and 
various temperatures obtained with SCO-RCG and STA codes.}
  \label{table4}
\end{table}

For completeness, we also listed in Table I the average ionization 
values obtained from our Average-Atom, STA, and SCO-RCG codes 
at a density of 0.1 g/cm$^3$ at five different temperatures. 
At this density, across the temperature range, the table shows 
STA values are consistently and slightly higher than the values 
obtained from the SCO-RCG calculations, indicating the predictions 
of a slightly higher electronic densities. 

\section{Conclusions}
The L-shell emission spectrum of germanium from a disc of 
0.1-$\mu$m-thick Ti/Ge mixture (sandwiched in a plastic) 
irradiated by a high-intensity (i.e., 10$^{17}$$-$10$^{19}$ W/cm$^2$), 
0.5 psec laser-pulse has been measured. The conditions of the plasma 
were inferred from simulations performed using both the LTE 
and non-LTE opacity codes. The non-LTE collisional 
radiative FLYCHK code estimated the plasmas to be at a density of 
1.5 $\pm$ 0.5 g/cm$^3$ and an electron temperature of
800 eV $\pm$ 100 eV. On the other hand, three LTE opacity 
codes equipped with various levels of sophistication in atomic 
physics models found the plasmas to be at the same density 
as FLYCHK but at a lower electron temperature of 600 eV $\pm$ 60 eV. 

Motivated by the work of Hoarty and Harris {\it et al}, 
we used a combination of the radiation-hydrodynamics
FastRad3D and STA opacity codes to study their emission data. 
To model the data, first, we performed the FastRad3D calculations, 
in one-dimension, to obtain the time-dependent plasma density 
and temperature profiles of the target. We employed an ansatz 
in the calculations, which considered an instantaneous energy-deposition 
into the plasma electrons because the energy-deposition of the 
laser occurs in a time-scale much shorter than the hydrodynamics 
response time. The resultant profiles were subsequently used 
in the STA calculations to obtain the synthetic emission spectra 
and opacity of hot and dense germanium plasma.  

Comparing STA calculated emission spectrum with LTE-GRASP2K results, 
we obtained reasonable agreement. Our analysis of the partial spectral 
contributions to the time-integrated emission spectrum also illustrated 
the sample conditions strongly depend upon the plasma temporal variations 
in density and temperature. In comparison with the experimental data, 
STA showed sufficient line-structure to reproduce 
the major emission spectral features of 2p$-$3d, 2s$-$3p 
and 2p$-$4d transitions displayed by the experiment.

The LTE-model is commonly assumed to be 
valid for describing a hot and dense plasma as in LTE, the electrons 
and ions have high collisions frequency are in equilibrium. However, one 
cannot simply dismiss the possibility that the photons aren't 
in equilibrium with these particles. The disparity between the 
STA results and experimental data in the $\sim$1.7 keV regions 
may be an example of that $-$ the non-LTE effects as well as 
the presence of the spatial temperature and density variations 
in the plasma. The limitations of the STA to account for these effects 
and quantitatively modeling the experimental data is expected, 
and are underscoring the difficulty of the present approach. 
Thus, good agreement between the GRASP2K and 
experimental spectra could be fortuitous, as GRASP2K 
is an LTE model.

The STA computed opacity profiles, ionization population fractions 
and average ionization were all compared with the results obtained 
from the SCO-RCG calculations for various plasma temperatures 
and densities over the L- and M-shell spectral range. The results 
from the two theories agree very well. All said comparisons of STA 
results in the observed spectrum and opacity are considerably close 
while offering the advantage of computational speed for generating 
opacity-emissivity database of high-Z plasmas needed
for FastRad3D hydrodynamic simulations.

One final remark. Busquet's RADIOM model \cite{RADIOM1, RADIOM2, RADIOM3} 
for effective ionization temperature $T_z$ is an appealing and a simple 
way to introduce non-LTE effects in hydrocodes. Since the experimental 
emission spectra appears to be in nearing LTE condition, it is of interest 
and our plan to examine the validity of the RADIOM model by comparing 
its predictions with the Hoarty's experiment and results from collisional 
radiative FLYCHK calculations.   

\vspace*{0.1cm}
\centerline {\bf Acknowledgment}
\vspace*{0.2cm} 
The work is supported in part by the U.S. Department of Energy National Nuclear 
Security Administration. WJ, DB and JCP acknowledge the access to the cluster facility 
GMPCS of LUMAT (FR LUMAT 2764).

\end{document}